\documentclass[aps,prb,twocolumn,showpacs,superscriptaddress,floatfix]{revtex4-1}
\usepackage{graphicx,graphics}
\usepackage{dcolumn}
\usepackage{amsmath,amssymb,amsfonts}
\usepackage{latexsym,verbatim}
\usepackage{color}
\usepackage{ulem}
\usepackage{bm}
\usepackage[percent]{overpic}
\usepackage[breaklinks=true,colorlinks,citecolor=blue,linkcolor=blue,urlcolor=blue]{hyperref}
\newcommand{\br}{{\bm r}}
\newcommand{\bv}{{\bm v}}

\begin{document}
\title{Electron hydrodynamics dilemma: whirlpools or no whirlpools}
\author{Francesco M. D. Pellegrino}
\email{francesco.pellegrino@sns.it}
\affiliation{NEST, Istituto Nanoscienze-CNR and Scuola Normale Superiore, I-56126 Pisa,~Italy}
\author{Iacopo Torre}
\affiliation{NEST, Scuola Normale Superiore, I-56126 Pisa,~Italy}
\affiliation{Istituto Italiano di Tecnologia, Graphene Labs, Via Morego 30, I-16163 Genova,~Italy}
\author{Andre K. Geim}
\affiliation{School of Physics \& Astronomy, University of Manchester, Oxford Road, Manchester M13 9PL, United Kingdom}
\author{Marco Polini}
\affiliation{Istituto Italiano di Tecnologia, Graphene Labs, Via Morego 30, I-16163 Genova,~Italy}

\begin{abstract}
In highly viscous electron systems such as, for example, high quality graphene above liquid nitrogen temperature, a linear response to applied electric current becomes essentially nonlocal, which can give rise to a number of new and counterintuitive phenomena including negative nonlocal resistance and current whirlpools. It has also been shown that, although both effects originate from high electron viscosity, a negative voltage drop does not principally require current backflow. In this work, we study the role of geometry on viscous flow and show that confinement effects and relative positions of injector and collector contacts play a pivotal role in the occurrence of whirlpools. Certain geometries may exhibit backflow at arbitrarily small values of the electron viscosity, whereas others require a specific threshold value for whirlpools to emerge.
\end{abstract}

\maketitle

\begin{figure}[t]
\centering
\begin{overpic}[width=0.9\columnwidth]{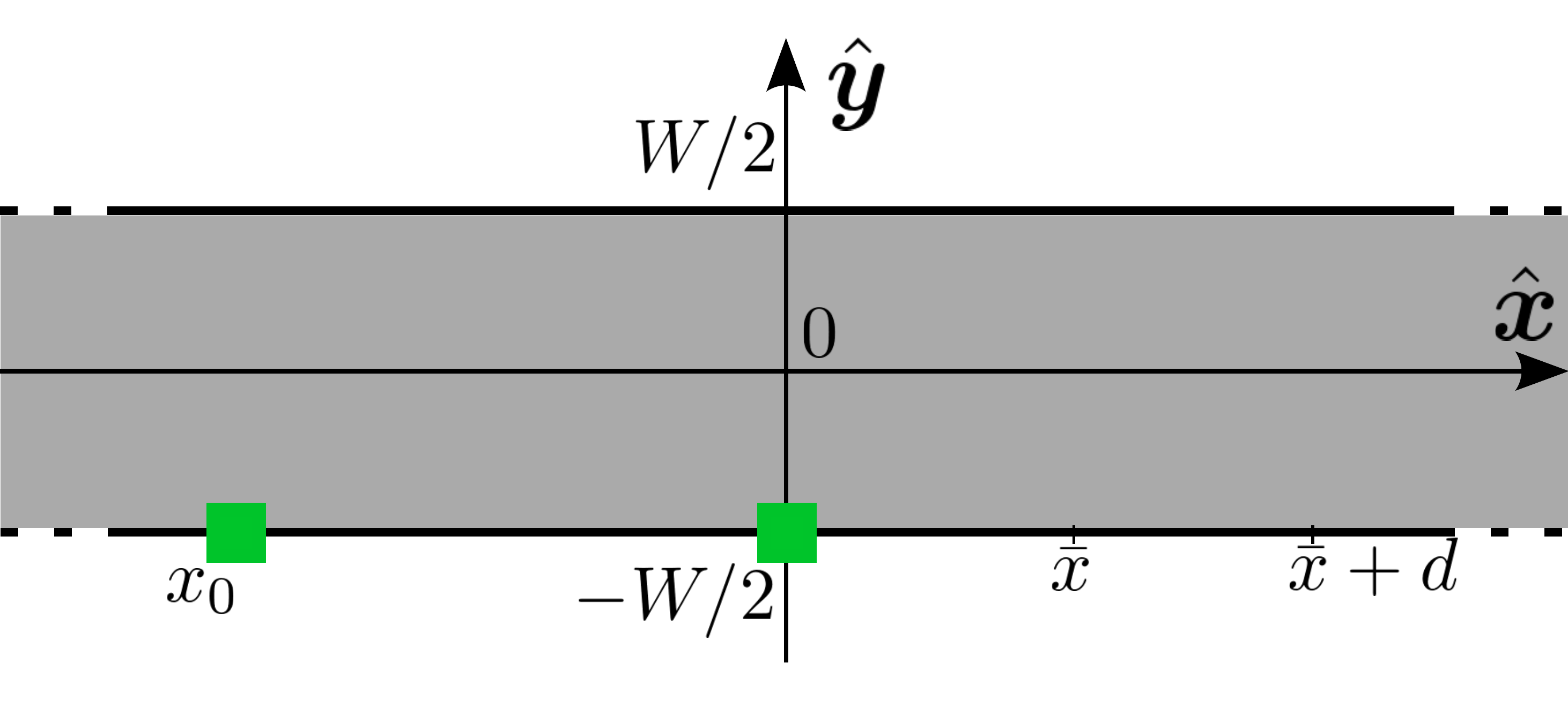}\put(2,35){(a)}
\end{overpic}\\
\begin{overpic}[width=0.9\columnwidth]{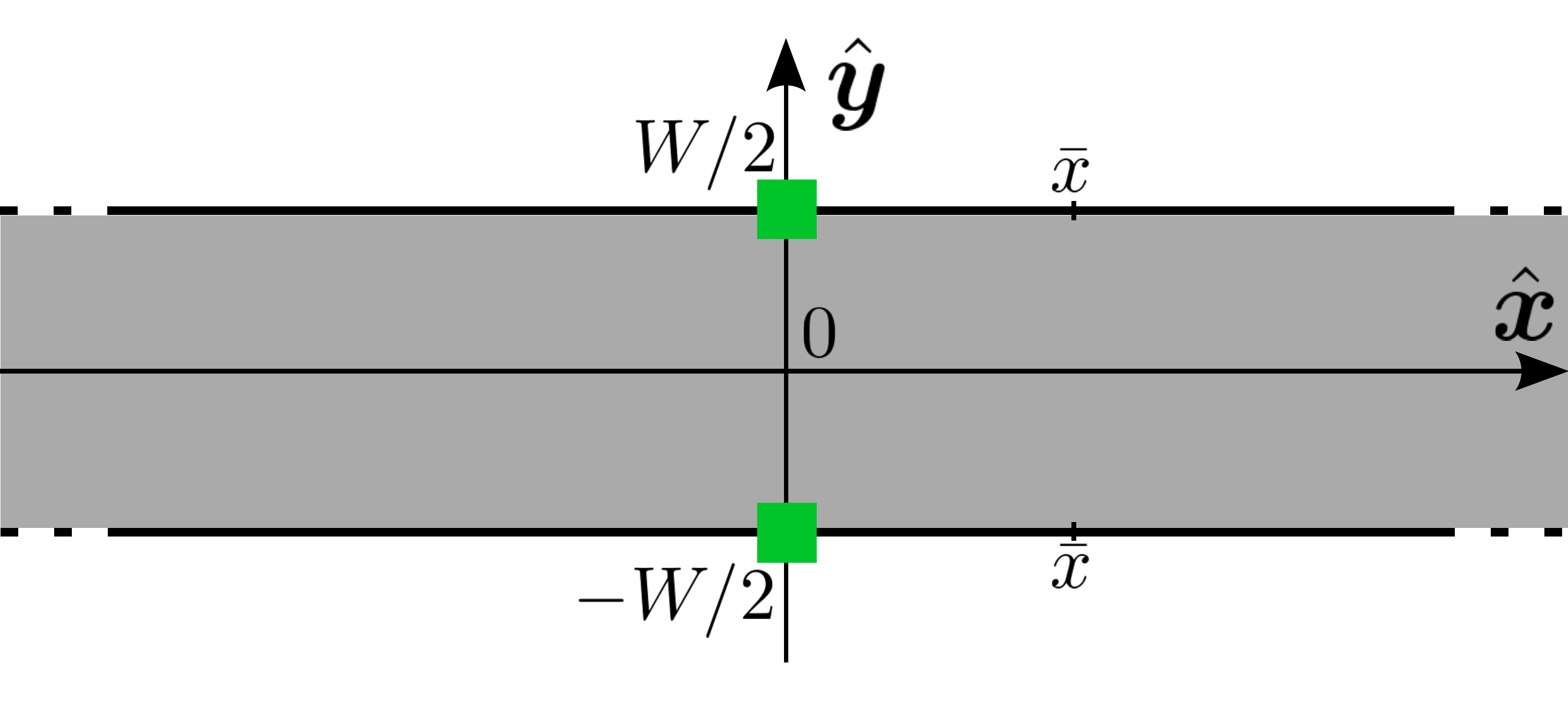}\put(2,35){(b)}
\end{overpic}\\
\begin{overpic}[width=0.9\columnwidth]{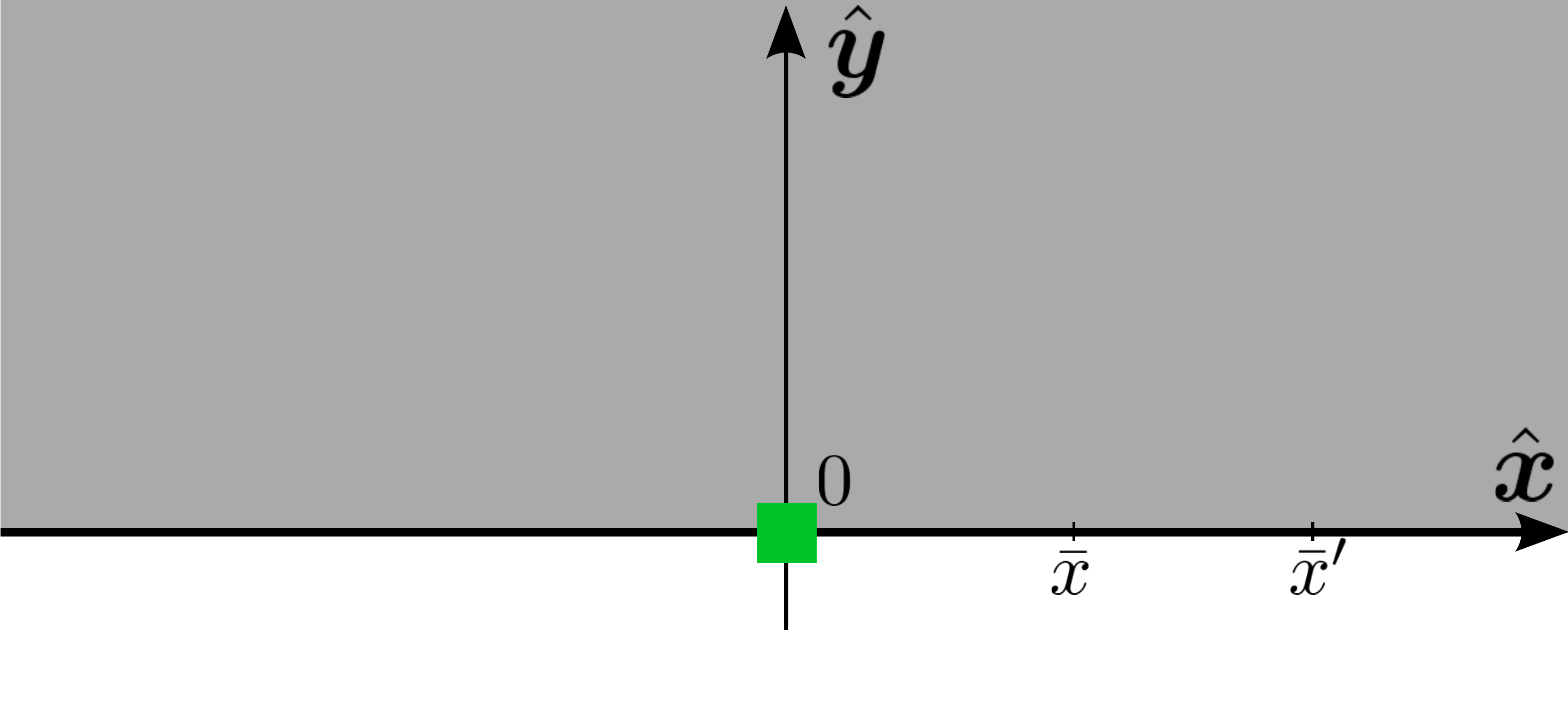}\put(2,48){(c)}
\end{overpic}
\caption{(Color online) A sketch of the nonlocal transport setups analyzed in this work. Both conductive channels (grey-shaded areas) in panels (a) and (b) have infinite length in the ${\hat {\bm x}}$ direction and finite width $W$ in the ${\hat {\bm y}}$ direction. The setup in panel (c) consists in a half-plane with a single edge located at $y=0$. Panel (a) illustrates the ``vicinity'' geometry~\cite{torre_prb_2015,bandurin_science_2016}. 
In this setup, current is injected into (extracted from) the green electrode located at $x=0$ ($x=x_{0}<0$) and $y=-W/2$. The nonlocal ``vicinity'' resistance is defined by $R_{\rm V} \equiv [\phi({\bar x},-W/2) - 
\phi({\bar x}+d,-W/2)]/I$, where $I$ is the injected current and $\phi(x,y)$ is the 2D electrostatic potential. 
For all practical purposes, we can take the limits $|x_{0}|, d \to +\infty$, which considerably simplify the final mathematical expression for $R_{\rm V}$. 
Panel (b) illustrates the LF geometry~\cite{levitov_naturephys_2016}. In this setup, current is injected into (extracted from) the green electrode located at $x=0$, $y = -W/2$ ($x=0$, $y = +W/2$). The nonlocal signal is defined by $R_{\rm LF} = [\phi(\bar{x},-W/2)-\phi(\bar{x},W/2)]/I$. Panel (c) illustrates the half-plane geometry. In this geometry, current is injected into a single electrode at the origin. The half-plane nonlocal resistance is defined as $R_{\rm HP} = [\phi(\bar{x},0)-\phi(\bar{x}',0)]/I$.\label{fig:setup}}
\end{figure}
\section{Introduction}
\label{sect:intro}

Hydrodynamics~\cite{landaufluidmechanics,falkovich_book} is a powerful non-perturbative theory for the  description of transport in materials where the mean free path $\ell_{\rm ee}$ for electron-electron (e-e) collisions happens to be much smaller than the sample size $W$ and the mean free path $\ell$ for momentum non-conserving collisions, i.e.~$\ell_{\rm ee} \ll \ell, W$. Despite the abundance of theoretical works~\cite{gurzhi_spu_1968,dyakonov_prl_1993,dyakonov_prb_1995,dyakonov_ieee_1996,govorov_prl_2004,muller_prb_2008,fritz_prb_2008,muller_prl_2009,bistritzer_prb_2009,andreev_prl_2011,mendoza_prl_2011,svintsov_jap_2012,mendoza_scirep_2013,tomadin_prb_2013,tomadin_prl_2014,torre_prb_2015_I,narozhny_prb_2015,briskot_prb_2015,torre_prb_2015,levitov_naturephys_2016,lucas_prb_2016}, clear-cut experimental evidence of hydrodynamic transport in the solid state has been lacking until recently, with the exception of early longitudinal transport experiments in electrostatically defined wires in the two-
dimensional (2D) electron gas in (Al,Ga)As heterostructures~\cite{molenkamp_sse_1994,dejong_prb_1995}. The latter reported the observation of negative differential resistance, which was interpreted as the Gurzhi effect~\cite{gurzhi_spu_1968} arising due to an increase in electron temperature due to current heating.

In graphene~\cite{geim_naturemater_2007}, hydrodynamic flow was originally predicted~\cite{muller_prb_2008,fritz_prb_2008,muller_prl_2009} to occur at the charge neutrality point (CNP), where thermally-excited electrons and holes undergo frequent collisions due to poorly-screened Coulomb interactions~\cite{kotov_rmp_2012}. In this regime, the authors of Ref.~\onlinecite{crossno_science_2016} have recently reported experimental evidence of the violation of the Wiedemann-Franz law, which is consistent with the occurrence of highly-frictional electron-hole flow.  

In the future, the strongly-interacting 2D electron-hole liquid in undoped graphene may enable investigations of solid-state nearly-perfect fluids~\cite{muller_prl_2009}, i.e.~fluids with very low values of the shear viscosity (in unit of the entropy density) and therefore minimal dissipation~\cite{NPFs}. At the CNP, however, carrier density inhomogeneities due to long-range disorder are unavoidable~\cite{dassarma_rmp_2011} and should be taken into account for a reliable description of the physics~\cite{lucas_prb_2016}. 

Microscopic calculations~\cite{li_prb_2013,polini_normale_2016,principi_prb_2016} suggest that also doped graphene sheets can display hydrodynamic behavior above liquid-nitrogen temperatures and for typical carrier concentrations. The reason is easy to understand. In the conventional Fermi-liquid regime, i.e.~for~$T \ll T_{\rm F} \equiv E_{\rm F}/\hbar$ where $E_{\rm F}$ is the Fermi energy, Pauli blocking is responsible for a very small rate of quasiparticle collisions
and  very long e-e mean free paths. In doped graphene~\cite{li_prb_2013,polini_normale_2016,principi_prb_2016}, $\ell_{\rm ee} \propto - 1/[T^2\ln(T)]$ for $T\ll T_{\rm F}$. As temperature increases, however, the Fermi surface ``softens", Pauli blocking is not as effective, and $\ell_{\rm ee}$ quickly decays, reaching a sub-micron size with an approximate power law $\ell_{\rm ee} \propto 1/T^{2}$. Furthermore, in 2D crystals where momentum-non-conserving collisions are dominated by acoustic phonon scattering, $\ell$ decays like $1/T$, thereby guaranteeing the existence of a temperature window where the hydrodynamic inequalities $\ell_{\rm ee} \ll \ell, W$ can be satisfied.

Doped graphene systems display very weak inhomogeneities due to the screening exerted on the long-range scattering sources by the electron liquid itself. Moreover, doped systems are characterized by large viscosities~\cite{principi_prb_2016,bandurin_science_2016} and values of $\ell_{\rm ee}$ that can be comparable to $\ell$, thereby offering an ideal platform to access a hydrodynamic regime in which quantum corrections to the Navier-Stokes equation are necessary, e.g.~in finite magnetic fields.

A recent experimental study~\cite{bandurin_science_2016} of ultra-clean single- and bi-layer graphene encapsulated between boron nitride crystals has indeed demonstrated that the 2D electron system in doped graphene displays hydrodynamic flow. For completeness, let us also mention recent reports on hydrodynamic transport in narrow quasi-2D channels of palladium cobaltate~\cite{moll_science_2016}. 

The authors of Ref.~\onlinecite{bandurin_science_2016} demonstrated that the nonlocal resistance in the so-called ``vicinity geometry''---Fig.~\ref{fig:setup}(a)---is negative in a carrier-density-dependent temperature window, as long as one is away from the CNP. This phenomenology was theoretically explained~\cite{torre_prb_2015,bandurin_science_2016} in terms of viscous contributions to the 2D electrostatic potential, which can be larger than canonical Ohmic contributions, therefore determining sign changes in nonlocal signals. In the geometry discussed in Refs.~\onlinecite{torre_prb_2015,bandurin_science_2016}, for example, nonlocal signals that are positive at low temperatures undergo two sign switches as temperature increases.
As we will see below, negative nonlocal resistance in the vicinity geometry comes together with {\it current whirlpools}. These are regions of the 2D steady-state current spatial pattern that display a vortex and backflow towards the current injector~\cite{bandurin_science_2016,torre_prb_2015}. 

A different nonlocal transport geometry was theoretically investigated by Levitov and Falkovich (LF)~\cite{levitov_naturephys_2016} and below referred to as the LF geometry. This is sketched in Fig.~\ref{fig:setup}(b). Theoretically, the LF geometry is highly symmetric and, as a consequence, when current whirlpools appear they do so along the longitudinal $\hat{\bm x}$-axis in the middle of the conductive channel ($y =0$). As a consequence, analytical calculations are simpler in the LF geometry than in the vicinity one. On the other hand, the latter is less prone to ballistic contributions that also result in negative resistance, which may severely obscure viscous effects~\cite{bandurin_science_2016}.  

\begin{figure}[t]
\centering
\includegraphics[width=0.9\columnwidth]{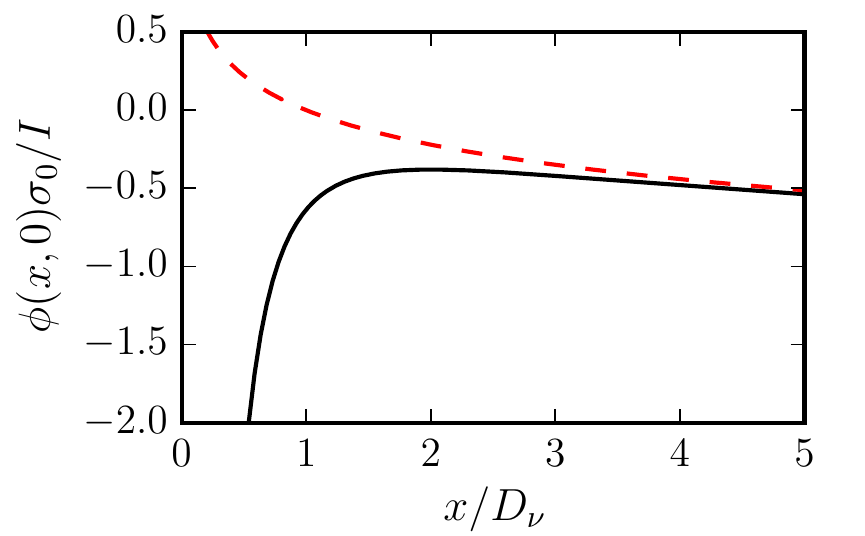}
\caption{(Color online) The solid line represents the dependence of the 2D electric potential $\phi(\br)$ on $x/D_{\nu}$ for a viscous 2D electron system confined to a half-plane. The potential is measured in units of $I/\sigma_{0}$ and is evaluated at the edge of the system, i.e.~at $y=0$. The Ohmic result in the absence of viscosity is also plotted (dashed line). We clearly see that viscosity introduces a region $\sim 2D_{\nu}$ near the injector where the 2D electrical potential is large and negative. \label{fig:pot_hp}}
\end{figure}

The third setup that will be analyzed in this work is sketched in Fig.~\ref{fig:setup}(c). It is a half-plane geometry with a single current injector, the simplest setup one can possibly imagine for the identification of viscosity-related features in nonlocal transport. The half-plane setup is conceptually very instructive, 
but of limited use to understand experiments that often involve finite-size devices (e.g., see Ref.~\onlinecite{bandurin_science_2016}). 

In this Article we present a comparative theoretical study of the three setups in Fig.~\ref{fig:setup}.
With the aid of {\it free-surface} boundary conditions~\cite{torre_prb_2015,bandurin_science_2016} on the linearized Navier-Stokes equation, we are able to find analytically the 2D electrostatic potential and steady-state distribution of currents in all the three geometries. We will emphasize the phenomenological features that these setups share but also some profound qualitative differences. As discussed in Refs.~\onlinecite{bandurin_science_2016,torre_prb_2015}, the use of free-surface boundary conditions is physically dictated by their compatibility with the measured~\cite{bandurin_science_2016} {\it monotonic} temperature dependence (i.e., no Gurzhi effect) of the ordinary longitudinal resistance $\rho_{xx}$ in the linear-response regime. On the other hand, the so-called {\it no-slip} boundary conditions~\cite{landaufluidmechanics} yield a non-monotonic temperature dependence of $\rho_{xx}$.

Before concluding, we would like to mention that other hydrodynamic models have been used earlier to discuss the behavior of 2D electron systems. 

Our Article is organized as follows. In Sect.~\ref{sect:theory} we review the theory of hydrodynamic transport in viscous 2D electron systems. In Sect.~\ref{sect:halfplane-geometry} we present the analytical solution of the problem in the case of the half-plane geometry. Similarly, in Sects.~\ref{sect:LF-geometry} and~\ref{sect:vicinity-geometry}, we present analytical solutions for the LF and vicinity geometries, respectively.
Finally, in Sect.~\ref{sect:conclusions} we summarize our principal findings and draw our main conclusions.

\section{Theory of hydrodynamic transport in viscous 2D electron systems}
\label{sect:theory}

In this Section we briefly review the theoretical approach that was introduced in Refs.~\onlinecite{bandurin_science_2016,torre_prb_2015} to study nonlocal transport in viscous 2D electron systems.

In the linear-response regime and under steady-state conditions, hydrodynamic transport in viscous 2D electron systems is governed by the continuity equation
\begin{equation}\label{eq:incompress}
\nabla \cdot \bv(\br)=0
\end{equation}
and the Navier-Stokes equation,
\begin{equation}\label{eq:ns1}
\frac{e}{m}  \nabla \phi (\br) + \nu \nabla^2 \bv (\br) - \frac{1}{\tau}\bv(\br)= 0~.
\end{equation}
Here, $-e$ is the electron charge, $m$ is the electron effective mass, $\bv(\br)$ is the fluid-element velocity field, $\phi(\br)$ is the 2D electrostatic potential, $\nu$ is the kinematic viscosity, and $\tau$ is a phenomenological transport time describing momentum-non-conserving collisions (e.g.~acoustic phonons). We emphasize that the continuity equation can be written as in Eq.~(\ref{eq:incompress})  since the 2D electron system behaves~\cite{bandurin_science_2016,torre_prb_2015} as an {\it incompressible} fluid to linear order in the drive current $I$. Beyond linear-response theory, the 2D electron system behaves as a compressible liquid. In this case, one needs to include in the set of hydrodynamic variables the local density $n({\bm r}) = {\bar n} +\delta n({\bm r})$, coupling the continuity equation and the Navier-Stokes equation with the three-dimensional Poisson equation~\cite{torre_prb_2015}. Here, ${\bar n}$ is the ground-state uniform electron/hole density.

\begin{figure}[t]
\centering
\includegraphics[width=0.9\columnwidth]{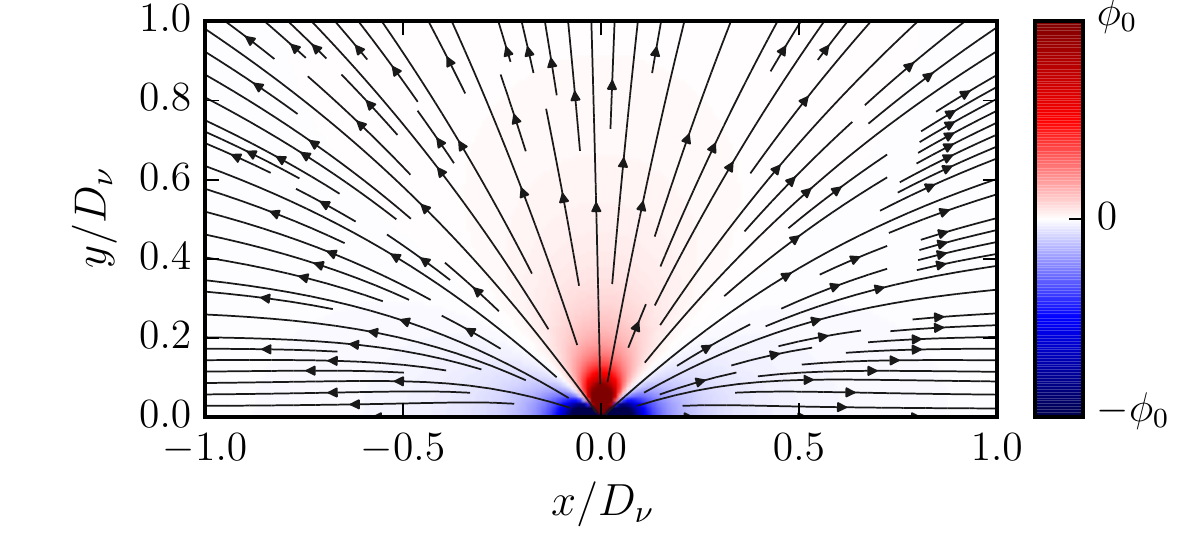}
\caption{(Color online) Nonlocal transport in a viscous 2D electron system confined to a half-plane geometry, as in Fig.~\ref{fig:setup}(c). The color map shows the 2D electric potential $\phi({\bm r})$ (in units of $\phi_0=100 I/\sigma_0$). The vector field represents the 2D charge current profile ${\bm J}({\bm r})$. Notice the absence of current whirlpools in this geometry. Asymptotically near the injector, we find ${\bm J}({\bm r}) \to 2I \sin^2(\theta)\br/(\pi r^2)$, where $\theta$ is the polar angle of $\br$. This result does not depend on the boundary conditions that are used to solve the problem, free-surface (this work and Refs.~\onlinecite{bandurin_science_2016,torre_prb_2015}) versus no-slip~\cite{levitov_naturephys_2016} boundary conditions. \label{fig:hp_maps}}
\end{figure}

Since all the setups in Fig.~\ref{fig:setup} are translationally-invariant in the ${\hat {\bm x}}$ direction, it is useful to introduce the Fourier transform with respect the spatial coordinate $x$:
\begin{equation}
\tilde{\phi}(k,y) = \int_{-\infty}^{+\infty}d x~e^{-i k x} \phi(\br)
\end{equation}
and
\begin{equation}
\tilde{\bv}(k,y) = \int_{-\infty}^{+\infty}d x~e^{-i k x} \bv(\br)~.
\end{equation}
The three coupled partial-differential equations (\ref{eq:incompress})-(\ref{eq:ns1}) can be combined into a $4\times4$ system of first-order ordinary differential equations:
\begin{equation}\label{eq:M-matrix}
\partial_y{\bm w}(k,y) = {\cal M}(k) {\bm w}(k,y)~,
\end{equation}
where ${\bm w}(k,y)$ is a four-component vector, ${\bm w}(k,y)=[k \tilde{v}_x(k,y),k \tilde{v}_y(k,y),\partial_y\tilde{v}_x(k,y),k^2  \sigma_0 \tilde{\phi}(k,y)/(e \bar{n})]^{\rm T}$, and
\begin{widetext}
\begin{equation}\label{eq:Mmatrix}
{\cal M}(k) = k \left(
 \begin{array}{cccc}
0 & 0 & 1  & 0  \\
-i & 0 & 0 & 0 \\
1+1/(k D_{\nu})^2 & 0 &0 & -i/(k D_{\nu})^2\\
0 &1 +(k D_{\nu})^2  &  i(k D_{\nu})^2 &0
\end{array}
\right)~,
\end{equation}
\end{widetext}
with $D_{\nu}\equiv \sqrt{\nu \tau}$ and $\sigma_{0} \equiv {\bar n} e^2 \tau/m$. The quantity $D_{\nu}$ represents the vorticity diffusion length~\cite{bandurin_science_2016,torre_prb_2015}, while $\sigma_{0}$ represents a Drude-like conductivity.

It can be easily checked that the matrix ${\cal M}(k)$ has four eigenvalues: $\lambda_{1,2}(k) = \pm 1$ and $\lambda_{3,4}(k) = \pm \sqrt{1+1/(k D_{\nu})^2 }$. The corresponding eigenvectors are: ${\bm w}_{1}(k) = (i,1,i,1)^{\rm T}$, ${\bm w}_2(k) = (i,-1,-i,1)^{\rm T}$, ${\bm w}_{3}(k) = \left(k/q,- i k^2/q^2,1,0 \right)^{\rm T}$, and ${\bm w}_{4}(k) = \left(-k/q,- i k^2/q^2,1,0 \right)^{\rm T}$, where we have introduced the shorthand
\begin{equation}\label{eq:qofk}
q \equiv q(k)=\sqrt{k^2+1/D_\nu^2}~.
\end{equation}
Eqs.~(\ref{eq:M-matrix})-(\ref{eq:qofk}) show that viscous transport is intrinsically nonlocal on the scale given by $D_{\nu}$.

The general solution of Eq.~(\ref{eq:M-matrix}) can be therefore written as a linear combination of exponentials of the form $\sum_{j=1}^4 a_j(k){\bm w}_j(k)\exp(\lambda_j k y)$, where ${\bm w}_j(k)$ and $\lambda_j(k)$  are eigenvectors and eigenvalues of the matrix ${\cal M}$, respectively. 
The four coefficients $a_j(k)$ can be determined from the enforcement of suitable boundary conditions (BCs).

\section{Half-plane geometry}
\label{sect:halfplane-geometry}

In the half-plane geometry, depicted in Fig.~\ref{fig:setup}(c), we consider a single current injector, which is described by the usual~\cite{abanin_science_2011} point-like BC for the component of the velocity field perpendicular to the edge:
\begin{equation}\label{eq:inj1} 
 v_y(x,y= 0)= - \frac{I}{e \bar{n}} \delta(x)~,
\end{equation}
where $I$ in the dc drive current. The solution of the viscous problem requires an additional BC on the tangential component of the velocity at the $y=0$ edge. Following Ref.~\onlinecite{torre_prb_2015}, one can work with a generic BC of the type
\begin{equation}\label{eq:tangBC1}
\left[\partial_y v_{x}(\br)+\partial_x v_y(\br)\right]_{y=0}= \frac{1}{\ell_{\rm b}} v_x(x,y=0)~, 
\end{equation}
where $\ell_{\rm b}$ is a boundary slip length~\cite{torre_prb_2015}. Finally, we also impose the following BCs at $y = +\infty$: $v_x(x,y \to +\infty )= 0$ and $v_y(x,y \to + \infty )= 0$. Note that the second term in square brackets in the left-hand side of Eq.~(\ref{eq:tangBC1}), i.e.~$\partial_x v_y(\br)$, is non-zero at the $y=0$ edge and must be retained. Indeed, inserting Eq.~(\ref{eq:inj1}) in Eq.~(\ref{eq:tangBC1}), we can rewrite the BC (\ref{eq:tangBC1}) more explicitly as
\begin{equation}\label{eq:tangBC1_explicit}
\left[\partial_y v_{x}(\br)\right]_{y=0} -\frac{I}{e {\bar n}}\delta^\prime(x)= \frac{1}{\ell_{\rm b}} v_x(x,y=0)~.
\end{equation}
In Fourier transform with respect to $x$, the BCs become
\begin{equation}\label{eq:ellb}
[\partial_y  \tilde{v}_{x}(k,y)+i k \tilde{v}_y(k,y) ] |_{y=0}=\frac{1}{\ell_{\rm b}} \tilde{v}_{x}(k,y=0)~,
\end{equation}
$\tilde{v}_y(k, y = 0) =- I/(e \bar{n})$, 
$\tilde{v}_x(k,y \to +\infty) = 0$, and 
$\tilde{v}_y(k,y \to +\infty) = 0$.

Imposing them we find the complete solution of the problem in Fourier transform with respect to $x$:
\begin{widetext}
\begin{equation}\label{eq:phi_halfplane_FT}
 \tilde{\phi}(k,y)= -\frac{I}{\sigma_0 } \frac{1}{|k|}
\frac{e^{-|k |y} \left[\ell_{\rm b} \left(k^2+q^2\right)+q\right]}{(|k|-q)[\ell_{\rm b} (|k|+q)+1]}~,
\end{equation}
\begin{equation}\label{eq:vx_halfplane_FT}
\tilde{v}_x(k,y)= - \frac{I}{e \bar{n}}\frac{i k}{|k|}
\Bigg\{
\frac{  \left[\ell_{\rm b} \left(k^2+q^2\right)+q\right]e^{-|k| y}}{(|k|-q) [\ell_{\rm b} (|k|+q)+1]}-
\frac{ q (2 |k| \ell_{\rm b}+1) e^{-q y}}{(|k|-q) [\ell_{\rm b} (|k|+q)+1]}
 \Bigg\}~,
\end{equation}
and
\begin{equation}\label{eq:vy_halfplane_FT}
\tilde{v}_y(k,y)= \frac{I}{e \bar{n}}
\Bigg\{
\frac{ \left[\ell_{\rm b} \left(k^2+q^2\right)+q\right]e^{-|k| y}}{(|k|-q) [\ell_{\rm b} (|k|+q)+1]}-
 \frac{|k| (2 |k| \ell_{\rm b}+1) e^{-q y}}{(|k|-q) [\ell_{\rm b} (|k|+q)+1]}
 \Bigg\}~.
\end{equation}
\end{widetext}

In the case of the {\it free-surface} BCs, which are obtained by taking the limit $\ell_{\rm b} \to +\infty$ in Eqs.~(\ref{eq:tangBC1}) and~(\ref{eq:ellb}),  the inverse  Fourier transforms of Eqs.~(\ref{eq:phi_halfplane_FT}), (\ref{eq:vx_halfplane_FT}), and (\ref{eq:vy_halfplane_FT}) can be calculated analytically. Simple mathematical manipulations allow us to find the electric
potential and the steady-state charge current for $\ell_{\rm b}\to +\infty$:
\begin{equation}\label{eq:hpphi}
\phi(\br)= -\frac{I}{\sigma_0 } (1-2D_\nu^2 \partial^2_x)  {\cal F}(\br) 
\end{equation}
and
\begin{equation}\label{eq:hpJ} 
{\bm J(\br)}\equiv-e  \bar{n} \bv(\br)=I \Big\{ \nabla {\cal F}(\br) + \nabla \times [\hat{\bm z} {\cal G}(D_{\nu}; \br) ]\Big\}~.
\end{equation}
In Eqs.~(\ref{eq:hpphi})-(\ref{eq:hpJ}) 
we have introduced the following auxiliary functions:
\begin{equation}\label{eq:Fhp}
{\cal F}(\br) =  \frac{1}{\pi} \ln(r/D_\nu)
\end{equation}
and
\begin{eqnarray}\label{eq:Ghp}
 {\cal G}(D_\nu; \br)&=& 2 D_\nu^2 \partial_x \partial_y \big[ 
   {\cal F}(\br) +\frac{1}{\pi}K_0(r/D_\nu) \big]~,
\end{eqnarray}
where $K_{0}(r/D_\nu)$ is the zeroth-order modified Bessel function of the second kind. 

Note that Eqs.~(\ref{eq:hpphi}) and~(\ref{eq:hpJ}) are manifestly universal, provided that one measures $x$ and $y$ in units of $D_{\nu}$, the potential in units of $I/\sigma_{0}$, and ${\bm J}$ in units of $I/D_{\nu}$. This stems, of course, from the fact that in the half-plane geometry there is one length scale, i.e.~the vorticity diffusion length $D_{\nu}$. 

In Eq.~(\ref{eq:hpphi}) we clearly see that the electric potential is the sum of an Ohmic contribution and a viscous one, which is proportional to $D^2_{\nu}$. Along the edge of the half-plane, the Ohmic result is positive definite, while the result in the presence of viscosity is large and negative: the viscous contribution to the potential dominates in the proximity of the current injector. Note that the Ohmic contributions to the potential and charge current density do {\it not} depend on $D_{\nu}$. Indeed, the Ohmic potential depends on $D_{\nu}$ only through a trivial constant, which has been introduced to make sure that the argument of the logarithm is dimensionless. Similarly, the Ohmic contribution to the current density does not depend on $D_{\nu}$, since the spatial derivative of a constant is zero.

Fig.~\ref{fig:pot_hp}  shows the 2D electric potential $\phi(\br)$ evaluated at the $y=0$ edge. In this figure, we only show $x>0$ since $\phi(-x,0)=\phi(x,0)$. Note that the electric potential is an increasing function of $x$ for $0<x \leq 2 D_{\nu}$. Defining the nonlocal voltage along the edge as
\begin{equation}\label{eq:VNLhp}
R_{\rm HP}(\bar{x})= \frac{\phi(\bar{x},0)-\phi(\bar{x}',0)}{I}~,
\end{equation}
we conclude that, in this ultra-simplified geometry, a clear signature of the role of viscosity in transport requires to probe the 2D electric potential in the close proximity of the injector, i.e.~for $\bar{x},\bar{x}' < 2 D_{\nu}$. 

We conclude this Section with two remarks on the steady-state charge current distribution pertaining the half-plane geometry:

(a) Fig.~\ref{fig:hp_maps} shows the universal spatial map of the 2D electric potential and the universal charge current streamlines in the half-plane geometry: independently of the value of $D_\nu$, {\it no current vortices and backflow occur in this geometry}.

(b) The current distribution ${\bm J}(\br)$ near the injector is {\it independent} of the BCs that are used. Indeed, for the case of free-surface BCs, expanding Eq.~(\ref{eq:hpJ}) near the current injector located at the origin, we find:
\begin{equation}\label{eq:asymptotics}
\lim_{r/D_{\nu} \to 0}{\bm J(\br)}= \frac{2 I \sin^2(\theta)}{\pi r^2}\br~,
\end{equation}
where $\theta$ is the polar angle of the vector $\br$. With no-slip BCs, i.e.~for $\ell_{\rm b}=0$, one finds exactly the same result.

The analytical solution of the problem in the half-plane geometry offers a situation in which negative nonlocal resistance near current injectors---Fig.~\ref{fig:pot_hp}---occurs in the absence of current whirlpools, i.e.~in the absence of backflow---Fig.~\ref{fig:hp_maps}. A natural question therefore arises: how general is this fact? Sects.~\ref{sect:LF-geometry} and~\ref{sect:vicinity-geometry} below answer this question.

\section{The LF geometry}
\label{sect:LF-geometry}

Here, we present analytical results for the setup~\cite{levitov_naturephys_2016} reported in Fig.~\ref{fig:setup}(b). 

In the LF geometry~\cite{levitov_naturephys_2016}, the BCs are:
\begin{equation}\label{eq:inj2} 
 v_y(x,y=\pm W/2) = - \frac{I}{e \bar{n}} \delta(x)
\end{equation}
and
\begin{equation}\label{eq:tangBC2}
\left[\partial_y v_{x}(\br)+\partial_x v_y(\br)\right] |_{y=\pm W/2}=\mp \frac{1}{\ell_{\rm b}} v_x(x,y=\pm W/2)~.
\end{equation}

Following the procedure outlined in Sects.~\ref{sect:theory}-\ref{sect:halfplane-geometry}, the solution in Fourier space for arbitrary boundary scattering length $\ell_{\rm b}$ reads as following:
\begin{widetext}
\begin{eqnarray}\label{eq:phigen}
 \tilde{\phi}(k,y) &=& 
\frac{I}{\sigma_0}
\sinh (k y) \big[ \ell_{\rm b} \left(k^2+q^2\right) \cosh \left(q W/2\right) +q \sinh \left(q W/2\right )\big]
/\big\{k\cosh \left(k W/2\right) \ \\
&\times&\big[\ell_{\rm b} \left(k^2-q^2\right) \cosh \left(q W/2\right) 
-q \sinh \left(q W/2\right)\big]+k^2 \sinh
   \left(k W/2\right) \cosh \left(q W/2\right)\big\}~, \nonumber
\end{eqnarray}
\begin{eqnarray}\label{eq:vxgen}
\tilde{v}_x(k,y) &=&  
-\frac{I}{e \bar{n}}  i  \big\{ q \sinh (q y) 
 \left[ 2 k \ell_{\rm b} \cosh \left(k W/2\right)+\sinh \left(k W/2\right)\right]
 - \sinh (k y) \big[q \sinh \left(q W/2\right)  \nonumber \\
 &+& \ell_{\rm b} \left(k^2+q^2\right) \cosh \left(q W/2\right)  \big] \big\}
/\big\{\cosh \left(k W/2\right) \big[\ell_{\rm b} \left(k^2-q^2\right) \cosh \left(q W/2\right) \nonumber \\
&-&q \sinh \left(q W/2\right)\big]+k \sinh
   \left(k W/2\right) \cosh \left(q W/2\right)\big\}~,
\end{eqnarray}
and
\begin{eqnarray}\label{eq:vygen}
\tilde{v}_y(k,y) &=&  
-\frac{I}{e \bar{n}}  \big\{ k \cosh (q y)\left[ 2 k \ell_{\rm b} \cosh \left(k W/2\right)+\sinh \left(k W/2\right)\right]
 - \cosh (k y)[q  \sinh \left(q W/2\right)\nonumber \\
   &+&\ell_{\rm b} \left(k^2+q^2\right) \cosh \left(q W/2\right)]
   \big\}
/\big\{\cosh \left(k W/2\right) \big[\ell_{\rm b} \left(k^2-q^2\right) \cosh \left(q W/2\right) \nonumber \\
&-&q \sinh \left(q W/2\right)\big]+k \sinh
   \left(k W/2\right) \cosh \left(q W/2\right)\big\}~.
\end{eqnarray}
\end{widetext}

Once again, the use of free-surface BCs, which are obtained by taking the limit $\ell_{\rm b} \to +\infty$,  allows us to calculate analytically the inverse Fourier transforms of Eqs.~(\ref{eq:phigen}), (\ref{eq:vxgen}) and (\ref{eq:vygen}). After straightforward mathematical manipulations, we find
\begin{eqnarray}\label{eq:potential_LF_geometry}
\phi(\br) &=& -\frac{I}{\sigma_0 } (1-2D_\nu^2 \partial^2_x) [F(x,y+W/2) \nonumber\\
&-& F(x,y-W/2)]
\end{eqnarray}
and
\begin{eqnarray}\label{eq:current_LF_geometry}
{\bm J}(\br) &=& I  \Big\{ \nabla [F(x,y+W/2)-F(x,y-W/2)] \nonumber\\
&+& \nabla \times \hat{\bm z} [G(D_\nu;x,y+W/2) \nonumber\\
&-& G(D_\nu;x,y-W/2)] \Big\}~,
\end{eqnarray}
where we have introduced the following auxiliary functions 
\begin{equation}\label{eq:auxF}
F(\br) =  \frac{1}{2\pi} \ln[\cosh(\pi x/W )-\cos(\pi y/W)]~,
\end{equation}
\begin{equation}\label{eq:auxG}
G(D_\nu;\br)= 2 D_\nu^2 [ \partial_x \partial_y  F(\br) + S(\br)]~,
\end{equation}
and
\begin{equation}
S({\bm r})\equiv \sum_{n=1}^{\infty} \sin\left(\frac{n \pi  y}{W}\right)  \frac{n \pi}{W^2}  {\rm sgn}(x)e^{- |x|\sqrt{(n \pi/W)^2 +1/D_\nu^2 }}~.
\end{equation}

In this geometry, the nonlocal resistance was defined as~\cite{levitov_naturephys_2016}
\begin{equation}\label{eq:non_local_LF_geometry}
R_{\rm LF}(\bar{x})\equiv  \frac{\phi(\bar{x},-W/2)-\phi(\bar{x},W/2)}{I}=\frac{2 \phi(\bar{x},-W/2)}{I}~.
\end{equation}
Replacing Eq.~(\ref{eq:potential_LF_geometry}) in Eq.~(\ref{eq:non_local_LF_geometry}) we find
\begin{eqnarray}\label{eq:VNL}
R_{\rm LF}(\bar{x}) &=& -\frac{1}{\sigma_0} 
\bigg \{
\frac{1}{\pi}\ln \left[\tanh^2 \left(\frac{\pi\bar{x}}{2W} \right) \right] \nonumber\\
&+&
4 \pi \left(\frac{D_{\nu}}{W}\right)^2 \frac{\cosh(\pi\bar{x}/W)}{\sinh^2(\pi\bar{x}/W)}   
\bigg\}~.
\end{eqnarray}

\begin{figure}[t]
\centering
\includegraphics[width=0.9\columnwidth]{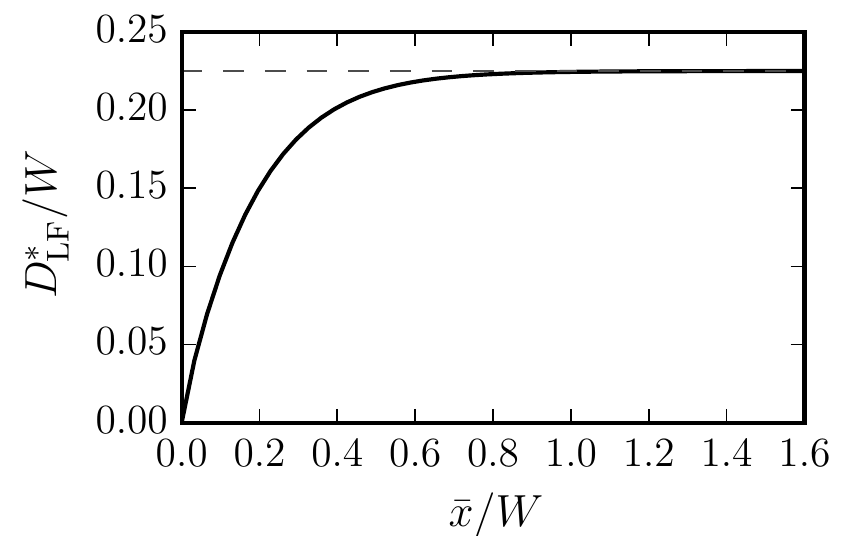}
\caption{The critical vorticity diffusion length $D^\ast_{\rm LF}({\bar x})$ (in units of $W$) defined in Eq.~(\ref{eq:Dnux}) is plotted as a function of $\bar{x}/W$. For ${\bar x}\gg W$, $D^\ast_{\rm LF}({\bar x}) \to W/(\sqrt{2}\pi)$ (horizontal dashed line).}
\label{fig:Dnu}
\end{figure}

We note that, for each lateral displacement $\bar{x}$ from the injector/collector electrodes in Fig.~\ref{fig:setup}(b), we can define the following critical vorticity diffusion length scale:
\begin{equation}\label{eq:Dnux}
D^\ast_{\rm LF}(\bar{x})=\frac{W}{2 \pi} \left\{-\frac{\displaystyle \sinh^2\left(\frac{\pi\bar{x}}{W} \right)}{\displaystyle \cosh\left(\frac{\pi\bar{x}}{W} \right)}
 \ln \left[\tanh^2 \left(\frac{\pi\bar{x}}{2W} \right) \right]
 \right\}^{1/2}~,
\end{equation}
which is such that $R_{\rm LF}(\bar{x})=0$. Fig.~\ref{fig:Dnu} shows $D^\ast_{\rm LF}$ as a function of $\bar{x}$. The physical meaning of the quantity $D^\ast_{\rm LF}(\bar{x})$ is the following. For $D_\nu> D^\ast_{\rm LF}(\bar{x})$, the nonlocal resistance $R_{\rm LF}(\bar{x})$ is {\it negative}.
Note that $D^\ast_{\rm LF}({\bar x}) \to 0$ for ${\bar x}\ll D_{\nu}$ and $D^\ast_{\rm LF}({\bar x}) \to W/(\sqrt{2}\pi)$ for ${\bar x}\gg W$. The first limit implies that, in the close proximity of the injector/collector electrodes, the nonlocal resistance $R_{\rm LF}(\bar{x})$ is negative for arbitrarily small values of the kinematic viscosity $\nu$.

\begin{figure}[t]
\centering
\includegraphics[width=0.9\columnwidth]{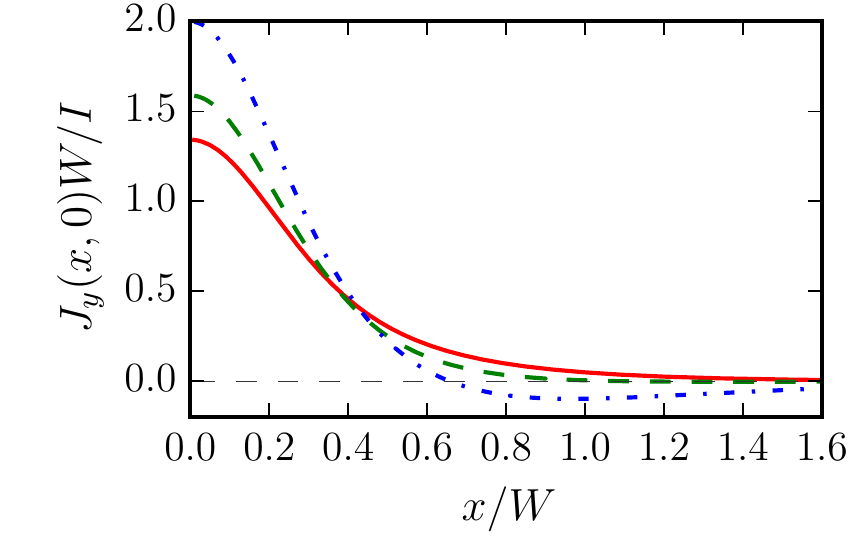}
\caption{(Color online) The quantity $J_y(x,0)$ (in units of $I/W$), calculated from Eq.~(\ref{eq:current_LF_geometry}), is plotted as a function of $x/W$. The solid line refers to $D_{\nu} = 0.15~W$, the dashed line to $D_{\nu}=0.25~W$, and the dash-dotted line to $D_\nu =10~W$.\label{fig:vy}}
\end{figure}

Now, the key question is: what about current whirlpools in this geometry? 
Without loss of generality, we can focus on the right side of the conductive channel, i.e.~for $x>0$. The setup in Fig.~\ref{fig:setup}(b) is clearly symmetric with respect to the inversion $x \to -x$.
Also, because of the symmetric location of the electrodes, the horizontal component of the current is identically zero along the $y=0$ axis, i.e.~$J_x(x,0)=0$. If a current vortex exists in this geometry, it must be centered on the $y=0$ axis. Fig.~\ref{fig:vy} shows the vertical component $J_y(x,0)$ of the current density as a function of $x$, for $y=0$. It is easy to show that $J_y(x,0)$ is positive at $x=0$, independently of the value of 
$D_{\nu}$. At large $x \gg W$ distances, on the other hand, one can approximate the current density along the $y=0$ axis as:
\begin{eqnarray}\label{eq:Jy_asy}
J_y(x\gg W,0) &\to&  \frac{2 I}{W} \bigg\{ [1 -2 \pi ^2 (D_\nu/W)^2] e^{-\pi  x/W} \nonumber\\
&+& 2 \pi ^2 (D_\nu/W)^2 e^{-x\sqrt{\frac{1}{D_\nu^2}+\frac{\pi ^2}{W^2}}}\bigg\}~.
\end{eqnarray}
Using Eq.~(\ref{eq:Jy_asy}), we find that $J_y(x\to +\infty,0) = 0^+$ for $D_{\nu}<W/(\sqrt{2} \pi)$, while 
$J_y(x\to +\infty,0) = 0^-$ for $D_\nu>W/(\sqrt{2} \pi)$.  We therefore conclude that 
$J_y(x,0)$ is positive for all the values of $x$ as long as $D_{\nu}<W/(\sqrt{2} \pi)$. In this geometry, current whirlpools do not exist for $D_\nu<W/(\sqrt{2} \pi)$. Plots of $J_y(x,0)$ for different values of $D_{\nu}$ are shown in Fig.~\ref{fig:vy}.

On the contrary, for $D_{\nu}>W/(\sqrt{2} \pi)$, there is a finite value of $x$, i.e.~$x_{\rm whirl}$, such that $J_y(x,0)<0$ for $x>x_{\rm whirl}$. This means that, for $D_\nu>W/(\sqrt{2} \pi)$, two current whirlpools appear in the LF geometry at positions $(\pm x_{\rm whirl},0)$. In particular, in the limit of a very large viscosity, i.e.~for $D_\nu \gg W$, one can write a closed-form expression for the current density. Indeed, in this limit,
the auxiliary function $G(D_\nu;\br)$ in Eq.~(\ref{eq:auxG}) tends to the following expression
\begin{equation}\label{eq:G_highvisc}
G(D_\nu \gg W;\br) =  - \frac{(x/W) \sin(\pi y/W)}{2[\cosh(\pi x/W)-\cos(\pi y/W)]}~.
\end{equation}
In this limit, $x_{\rm whirl}$ is the root of the transcendental equation $\pi x_{\rm whirl}\tanh(\pi x_{\rm whirl}/W)/W = 2$, yielding $x_{\rm whirl} \approx 0.66W$.

\begin{figure}[t]
\centering
\begin{overpic}[width=1.0\columnwidth]{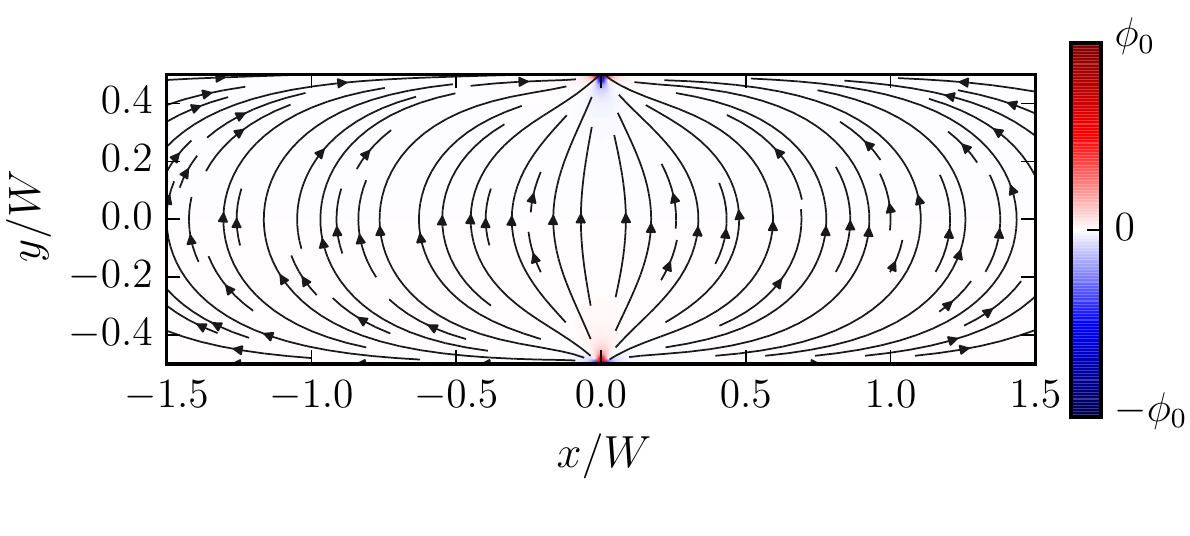}\put(1,40){(a)}\end{overpic}\\
\begin{overpic}[width=1.0\columnwidth]{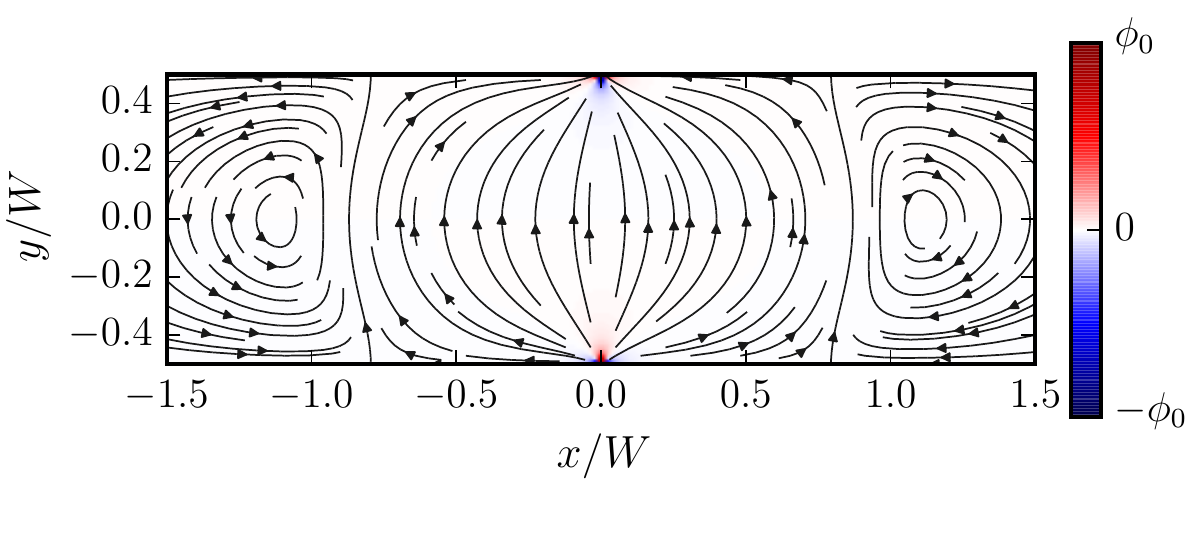}\put(1,40){(b)}\end{overpic}\\
\begin{overpic}[width=1.0\columnwidth]{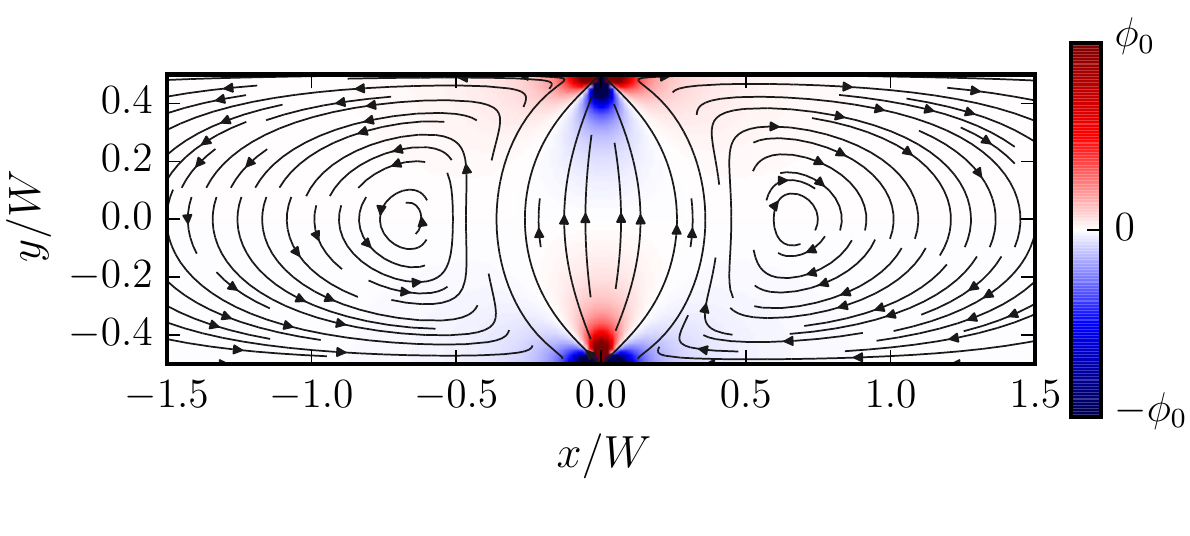}\put(1,40){(c)}\end{overpic}
\caption{(Color online) Nonlocal transport in the LF geometry---Fig.~\ref{fig:setup}(b). The color map denotes the spatial distribution of the 2D electric potential $\phi(\br)$ (in units of $\phi_0=100 I/\sigma_0$). The vector field denotes the charge current density ${\bm J}(\br)$. Panel (a): $D_{\nu} = 0.20 W$. Panel (b): $D_{\nu} = 0.25 W$. Panel (c): $D_\nu=W$. We clearly see current whirlpools in panels (b) and (c) because both values of $D_{\nu}$ that have been used to make these two plots are above the threshold value $D_{\nu} = W/(\sqrt{2}\pi)\simeq 0.225W$. \label{fig:levitov_maps}}
\end{figure}

In summary, in the LF geometry whirlpools emerge only above a threshold value of viscosity, i.e.~for $D_\nu \geq W/(\sqrt{2}\pi)$. 
At $D_{\nu} = W/(\sqrt{2}\pi)$, whirlpools form at infinity. For $D_\nu \gg W/(\sqrt{2}\pi)$, whirlpools approach the position $(\pm 0.66W,0)$. Typical results for 2D electric potential $\phi(\br)$ and charge current density ${\bm J}(\br)$ in this geometry are shown in Fig.~\ref{fig:levitov_maps}. 
For a highly viscous and clean electron system such as that in graphene, one can reach $D_{\nu}$ of  $\sim 0.3$-$0.4~{\rm \mu m}$~(Ref.~\onlinecite{bandurin_science_2016}), 
which necessitates devices with $W \lesssim 1.3$-$1.8~{\rm \mu m}$ to be able to create whirlpool currents. 

\section{The vicinity geometry}
\label{sect:vicinity-geometry}
\begin{figure}[t]
\centering
\includegraphics[width=0.8\columnwidth]{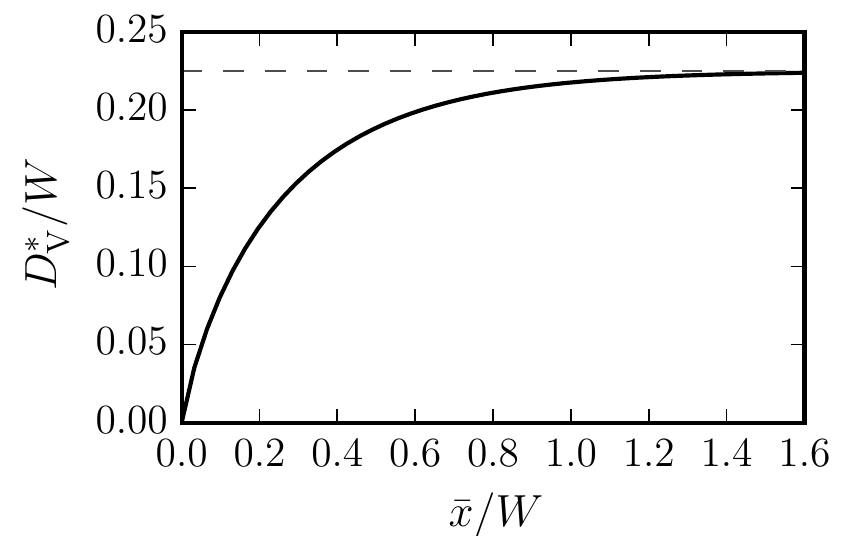}
\caption{(Color online) The critical vorticity diffusion length $D^\ast_{\rm V}({\bar x})$ (in units of $W$) defined in Eq.~(\ref{eq:Dnux_torre}) is plotted as a function of $\bar{x}/W$. For ${\bar x}\gg W$, $D^\ast_{\rm V}({\bar x}) \to W/(\sqrt{2}\pi)$ (horizontal dashed line).\label{fig:Dnu_torre}}
\end{figure}
\begin{figure}[t]
\centering
\includegraphics[width=0.8\columnwidth]{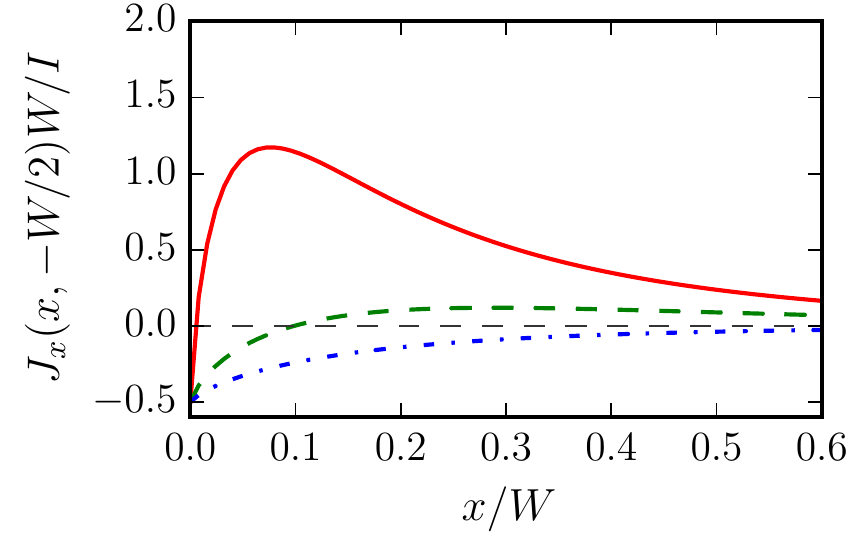}
\caption{(Color online) The quantity $J_x(x,-W/2)$ (in units of $I/W$), calculated from Eq.~(\ref{eq:current_V_geometry}), is plotted as a function of $x/W$. The solid line refers to $D_{\nu} = 0.05~W$, the dashed line to $D_{\nu}=0.15~W$, and the dash-dotted line to $D_\nu =0.25~W$.\label{fig:vxedge}}
\end{figure}

In this Section we present analytical results for the vicinity setup~\cite{torre_prb_2015,bandurin_science_2016} in Fig.~\ref{fig:setup}(a).
 
In this geometry, the BCs read as following
\begin{equation}\label{eq:inj3_first} 
v_y(x,y=+ W/2)=0~,
\end{equation}
\begin{equation}\label{eq:inj3_second}
v_y(x,y=- W/2)=  -\frac{I}{e \bar{n}}[\delta(x)-\delta(x-x_0)]~,
\end{equation}
while the free-surface BC on the tangential component of the fluid-element velocity reduces to
\begin{equation}\label{eq:tangFS}
\left[\partial_y v_{x}(\br)+\partial_x v_y(\br)\right] |_{y=\pm W/2}=0~. 
\end{equation}

Repeating the same algebraic steps outlined in the previous Sections, we find that the electric potential and charge current distribution in this geometry can be written as:
\begin{eqnarray}\label{eq:potential_V_geometry}
\phi(\br) &=& -\frac{I}{\sigma_0 } (1-2D_\nu^2 \partial^2_x) [F(x,y+W/2) \nonumber \\
&-& F(x-x_0,y+W/2)]
\end{eqnarray}
and
\begin{eqnarray}\label{eq:current_V_geometry}
 {\bm J}(\br) &=& I  \Big\{ \nabla [F(x,y+W/2)-F(x-x_0,y+W/2)] \nonumber\\
 &+& \nabla \times \hat{\bm z} [G(D_\nu;x,y+W/2) \nonumber \\
 &-& G(D_\nu;x-x_0,y+W/2)] \Big\}~,
\end{eqnarray}
where the auxiliary function $F(\br)$ and $G(D_\nu;\br)$ have been defined in Eqs.~(\ref{eq:auxF}) and~(\ref{eq:auxG}), respectively.

The nonlocal vicinity voltage can be defined as
\begin{equation}\label{eq:VNL_torre}
R_{\rm V}(\bar{x})\equiv \frac{\phi(\bar{x},-W/2)-\phi(\bar{x}+d,-W/2)}{I}~.
\end{equation}
The expression of the vicinity resistance notably simplifies in the limit $x_0 \to - \infty$ and $d \to +\infty$: taking these limits we find
\begin{eqnarray}
R_{\rm V}(\bar{x}) &=& -\frac{1}{2\sigma_0}
\Bigg\{
\frac{1}{\pi}\ln \left[4 \sinh^2 \left( \frac{\pi {\bar x}}{2 W} \right) \right] -\frac{{\bar x}}{W} \nonumber\\
 &+& \pi\left(\frac{D_{\nu}}{W}\right)^2 \frac{1}{\displaystyle \sinh^2 \left(\pi {\bar x}/(2 W) \right)}\Bigg\}~.
\end{eqnarray}

Similarly to what was done in Sect.~\ref{sect:LF-geometry}, we can define a critical vorticity diffusion length scale $D^\ast_{\rm V}(\bar{x})$ as following:
\begin{eqnarray}\label{eq:Dnux_torre}
 D^\ast_{\rm V}(\bar{x}) &\equiv&  W \sinh \left( \frac{\pi \bar{x}}{2 W} \right) \nonumber\\
 &\times&
 \bigg\{\frac{\bar{x}}{\pi W}-\frac{1}{\pi^2}\ln \left[4 \sinh^2 \left( \frac{\pi \bar{x}}{2 W} \right) \right]  \bigg\}^{1/2}~.
\end{eqnarray}
For $D_\nu> D^\ast_{\rm V}(\bar{x})$ the vicinity resistance $R_{\rm NL}(\bar{x})$ is negative. 
Fig.~\ref{fig:Dnu_torre} illustrates the functional dependence of $D^\ast_{\rm V}(\bar{x})$ on $\bar{x}$. As in the case of $D^\ast_{\rm LF}(\bar{x})$, $D^\ast_{\rm V}({\bar x})$ tends to the asymptotic value $W/(\sqrt{2}\pi)$ for $\bar{x}\gg W$.

Unlike the LF geometry, the vicinity one exhibits a more direct relation between negative nonlocal voltage and current whirlpools. 
In the proximity of the current injector, i.e.~for $x \ll D_{\nu},W$ and $y\to -W/2$, and in polar coordinates, the current density (\ref{eq:current_V_geometry}) behaves like
\begin{equation}\label{eq:Jinj}
{\bm J}(\br) \to I \left[ -\frac{1}{2 W}\hat{\bm x} + \frac{2 \sin^2(\theta)}{\pi r^2}\br\right]~,
\end{equation}
where we have used the asymptotic expansion (\ref{eq:asymptotics}) for the half-plane geometry. In Eq.~(\ref{eq:Jinj}) we have taken the origin of the polar plane to lie at $(0,-W/2)$. Note the presence of the first term in the square brackets in Eq.~(\ref{eq:Jinj}), i.e.~$-I/(2W)$, which is due to the collector at $x_0 \to -\infty$. This term has crucial implications on the occurrence of whirlpools in the vicinity geometry~\cite{torre_prb_2015,bandurin_science_2016}. Indeed, from the BC (\ref{eq:inj3_second}), we see that $J_y(x,-W/2)=0$ for $x>0$. 
Eq.~(\ref{eq:Jinj}) implies that $J_x(0,-W/2)=-I/(2 W)<0$, independently of the value of $D_{\nu}$.
{\it This implies that in the vicinity geometry there is always backflow in the proximity of the injector, independently of the value of $D_{\nu}$.} 

As we now proceed to demonstrate, the precise value of $D_{\nu}$ sets only the spatial extension of the current whirlpool. 
At large lateral separations from the injector, one can approximate the current density (\ref{eq:current_V_geometry}) along the bottom edge as
\begin{eqnarray}
J_x(x\gg W,-W/2) &\to& \frac{I}{W} \Big\{ [1 -2 \pi ^2 (D_\nu/W)^2] e^{-\pi  x/W} \nonumber\\
&+& 2 \pi ^2 (D_\nu/W)^2 e^{-x \sqrt{\frac{1}{D_\nu^2}+\frac{\pi ^2}{W^2}}}\Big\}~.
\end{eqnarray}
\begin{figure}[t]
\centering
\begin{overpic}[width=1.0\columnwidth]{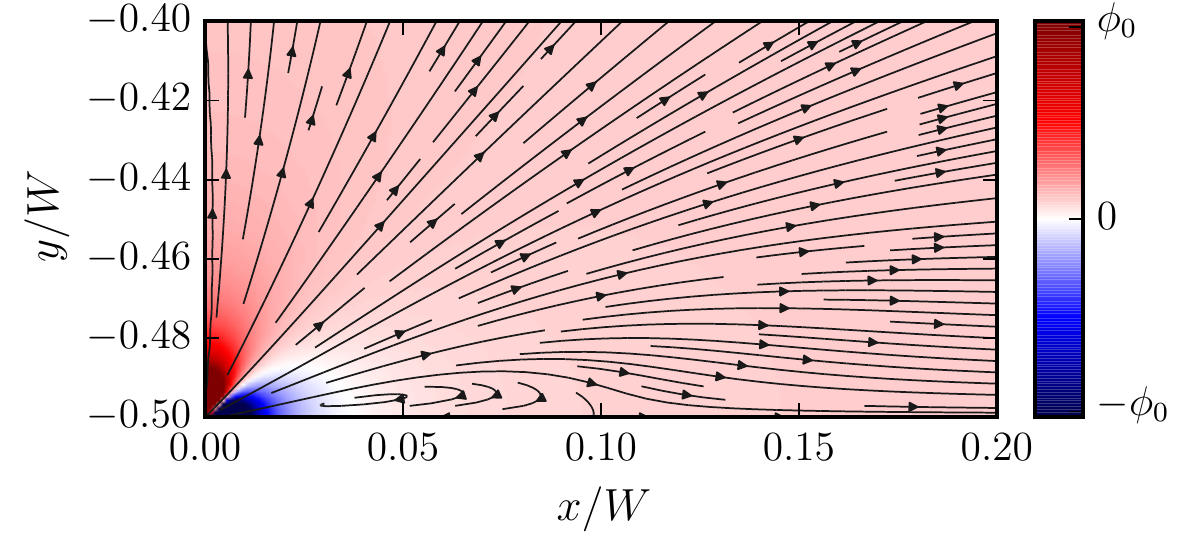}\put(1,40){(a)}\end{overpic}\\
\begin{overpic}[width=1.0\columnwidth]{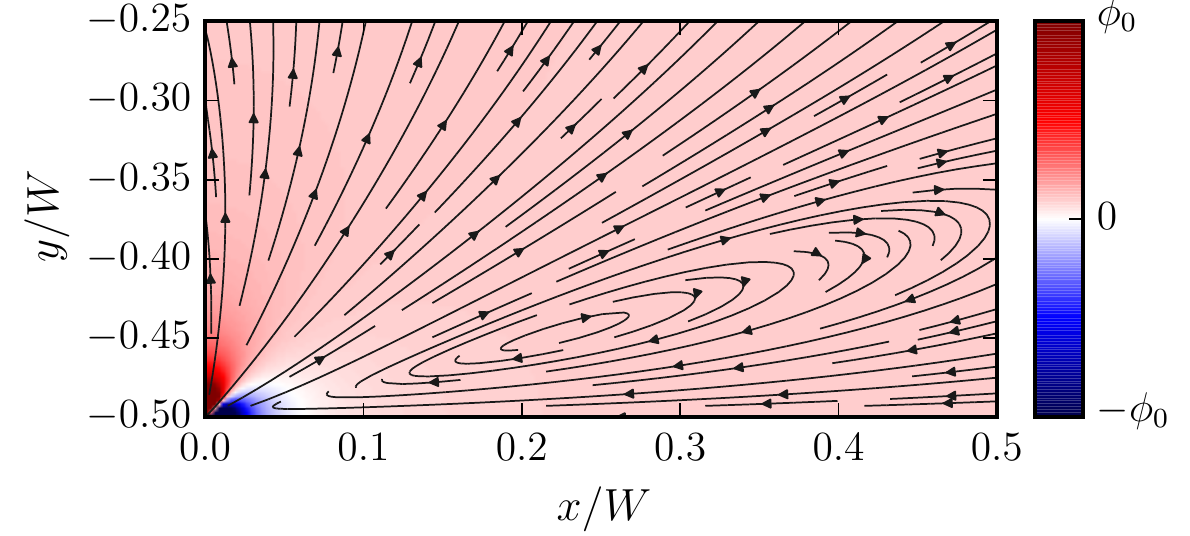}\put(1,40){(b)}\end{overpic}
\caption{(Color online)  Nonlocal transport in the vicinity geometry---Fig.~\ref{fig:setup}(a). The color map denotes the spatial distribution of the 2D electric potential $\phi(\br)$ (in units of $\phi_0=100 I/\sigma_0$). The vector field denotes the charge current density ${\bm J}(\br)$. Data in this plot refer to the spatial region $x>0$ in Fig.~\ref{fig:setup}(a). Panel (a): $D_\nu=0.15 W$. Panel (b) $D_\nu=0.25 W$. While backflow is present in both panels, the precise value of $D_{\nu}$ sets the spatial extension of current whirlpools.\label{fig:torre_maps}}
\end{figure}
Using the previous result, we find that $J_x(x \gg W,-W/2) = 0^+$ for $D_{\nu}<W/(\sqrt{2} \pi)$, while $J_x(x\gg W,-W/2) = 0^-$ for $D_\nu>W/(\sqrt{2} \pi)$.  This implies that $J_x(x,-W/2)$ is negative for {\it all} values of $x>0$ for $D_\nu>W/(\sqrt{2} \pi)$.
This is clearly seen in Fig.~\ref{fig:vxedge} for $D_\nu=0.25~W$ (dash-dotted line). On the contrary, for $D_\nu<W/(\sqrt{2} \pi)$, $J_x(x,-W/2)$ is negative in a {\it finite} range of values of $x>0$, as one can see in Fig.~\ref{fig:vxedge} for $D_\nu=0.05 W$ (solid line) and $D_\nu=0.15 W$ (dashed line). 

In Fig.~\ref{fig:torre_maps} we show that, independently of the value of $D_{\nu}$, viscosity induces a vortex to the right of the current injector.  For $D_\nu<W/(\sqrt{2} \pi)$, the vortex is ``localized'' in an increasingly smaller region in the close proximity of the current injector, as shown in Fig.~\ref{fig:torre_maps}(a), while for $D_\nu>W/(\sqrt{2} \pi)$ the vortex spreads out in space far away from the location of the current injector, as in Fig.~\ref{fig:torre_maps}(b).

In the experiments~\cite{bandurin_science_2016},  devices with $W$ ranging from $1.5$ to $4~{\rm \mu m}$ were employed which, for $D_{\nu} \approx 0.4~{\rm \mu m}$, yields $D_{\nu}/W \approx 0.27$ to $0.1$, respectively. For a vicinity contact placed at a distance of $1~{\rm \mu m}$, we have checked numerically that {\it backflow at the contact} is expected if $W \gtrsim 1.8~{\rm \mu m}$. 
In reality, however, this condition is softened by the fact that both injector and detector contacts had a finite (relatively large) width of $\approx 0.3~{\rm \mu m}$, which should allow backflow at a nominal distance to the injector larger than $2~{\rm \mu m}$. Nonetheless, even the device with $W = 4~{\rm \mu m}$ exhibited negative vicinity resistance, in agreement with the fact that the latter is a necessary but not sufficient condition for the existence of backflow at the vicinity contact.

\section{Summary and conclusions}
\label{sect:conclusions}

In this work we have studied the role of geometric effects in two-dimensional solid-state hydrodynamic transport. We have been able to demonstrate that they play a crucial role in the establishment of so-called current whirlpools~\cite{bandurin_science_2016,torre_prb_2015}. 

The half-plane geometry---sketched in Fig.~\ref{fig:setup}(c)---hosts negative nonlocal resistances due to viscosity but no current whirlpools. 

The geometry analyzed in Ref.~\onlinecite{levitov_naturephys_2016}, which is depicted in Fig.~\ref{fig:setup}(b), allows the formation of current whirlpools only if the electron liquid viscosity, at a given carrier density and temperature, overcomes a threshold value, i.e.~$D_{\nu} >W/(\sqrt{2}\pi)$ or, more explicitly, $\nu> W^2/(2\pi^2\tau)$.  

In contrast to the above two geometries, the vicinity geometry introduced in Refs.~\onlinecite{bandurin_science_2016,torre_prb_2015} and sketched in Fig.~\ref{fig:setup}(a) exhibits backflow {\it near} the injector electrode for arbitrarily small values of $D_{\nu}$. The value of $D_{\nu}$ affects the spatial extent of current whirlpools, as shown in Fig.~\ref{fig:torre_maps}.  
To detect current backflow in this geometry, either a local probe should be in the immediate vicinity of the injector or the width $W$ of the conductive channel should be chosen sufficiently small. For the case of graphene with its typical vorticity diffusion length $\approx 0.3$-$0.4~{\rm \mu m}$ and a distance of $1~{\rm \mu m}$ between a narrow probe and a current injector, $W$ should be $< 1.5$-$2~{\rm \mu m}$.

We hope that this work helps clarifying the subtle connection between backflow and negative nonlocal resistances due to viscosity in 2D electron liquids. We also hope that it will spark experimental quests of current whirlpools based on scanning probe potentiometry and magnetometry, as suggested in Ref.~\onlinecite{torre_prb_2015}.

\acknowledgements
Free software (www.gnu.org, www.python.org) was used. 
This work was supported by Fondazione Istituto Italiano di Tecnologia, Centro Siciliano di Fisica Nucleare e Struttura della Materia (CSFNSM), and, only in part, by the European Union's Horizon 2020 research and innovation programme under grant agreement No.~696656 ``GrapheneCore1''. We thank A.~Tomadin for fruitful discussions.

\end{document}